\documentclass[aps,prb,twocolumn,nopacs,floats,superscriptaddress]{revtex4-2}
\usepackage{graphicx}
\usepackage{dcolumn}
\usepackage{bm}
\usepackage{amsmath}
\usepackage{color}
\UseRawInputEncoding

\definecolor{cardinal}{rgb}{0.6,0,0}
\definecolor{darkgreen}{rgb}{0,0.4,0}
\definecolor{golden}{rgb}{0.92, 0.7, 0}
\definecolor{midnight}{rgb}{0, 0, 0.5}
\definecolor{darkblue}{rgb}{0, 0, 0.7}

\def\he4{$^4$He}
\def\hel3{$^3$He}
\def\Am3{\AA$^{-3}$}
\def\beq{\begin{equation}}
\def\eeq{\end{equation}}

\newcommand{\be}{\begin{equation}}
\newcommand{\ee}{\end{equation}}
\newcommand{\bea}{\begin{eqnarray}}
\newcommand{\eea}{\end{eqnarray}}
\newcommand{\bse}{\begin{subequations}}
\newcommand{\ese}{\end{subequations}}

\begin{document}
\author{Adham Alkady}
\affiliation{Department of Physics \& Astronomy, College of Staten Island and the Graduate Center of
CUNY, Staten Island, NY 10314}

\author{Victor Fleurov}
\affiliation{Raymond and Beverly Sackler Faculty of Exact Sciences,
School of Physics and Astronomy, Tel-Aviv University - Tel-Aviv 69978, Israel}

\author{Anatoly Kuklov}
\affiliation{Department of Physics \& Astronomy, College of Staten Island and the Graduate Center of
CUNY, Staten Island, NY 10314}


\title{Phonon-induced modification of polaritonic Rabi oscillations in the presence of the dark excitonic condensate}

\begin{abstract}
Optically inactive (dark) intervalley momentum-forbidden excitons are characterized by relatively long life time, and therefore are desirable candidates for realizing collective excitonic phases. However, testing their coherence by light directly is impossible.  Here we propose a method for detecting a dark excitonic condensate. It relies on the interaction between excitons and phonons responsible for the interconversion between bright and dark excitons. As long as the dark condensate forms, the Rabi oscillations between photons and bright excitons can become strongly modified, and can be viewed as the photon-exciton-phonon polaritonic effect.
The multi-component nature of the dark condensate consistent with the point-group symmetry is taken into account in the limit of weak phonon-exciton interaction. A perspective for the case of the strong interaction leading to the polaronic effect is discussed. 
\end{abstract}

\maketitle

\section{Introduction}

In contrast to the bright excitons, dark excitons cannot be created directly by light. Two typical examples are i) momentum forbidden (intervalley) excitons, and ii) excitons requiring spin flips which cannot be induced by photons.  A detailed account of the excitonic properties with the focus on i) and ii) in transition metal dichalcogenides (TMDs) is presented in Ref.\cite{BEDE}. Here we will be mainly discussing the first case, that is, the excitons formed by electrons and holes located in different band valleys shifted with respect to each other by a momentum ${\bf q}_0$ which is much larger than that of typical photons. As a specific example, we refer to the excitonic transition $\Gamma \Lambda'$ shown in Fig.1(a) of Ref.\cite{Gamma}  (a hole is at the $\Gamma$-point of the valence band and an electron is in the conduction band at the valley along the $\Lambda '$ line toward the $K'$ point) with its $N_v=6$ replicas connected by the hexagonal point group symmetry. As a result of this shift, a dark exciton carries finite momentum in its lowest energy state. Accordingly, the condensation of these excitons implies a formation of the quantum condensate characterized by finite momentum. At this point it is worth mentioning that a crystal group symmetry implies   that there can be several such identical condensates distinguished by its own valley momentum ${\bf q}_\alpha$ with $\alpha=1,2,....N_v$ and
$|{\bf q}_\alpha|= q_0$. Detecting such multi-component condensates and being able to study their properties is of great fundamental importance. Our proposal is a first step toward this goal. 

Dark exciton can normally persist much longer than the bright one \cite{long}, and this creates an attractive perspective for realizing  collective states of the excitonic ensembles such as  Bose-Einstein condensates \cite{Keldysh,Moskalenko} and other strongly interacting collective phases of light (see in Ref.\cite{Iacopo}). Various methods are used to detect such excitons (see, e.g., in Refs. \cite{WS2, TiO2}). 
One option stems from the two-exciton interconversion facilitated by the Coulomb interaction---when two bright ones transform into two dark and back.  Accordingly, the photoluminescence should demonstrate coherent oscillations as a signature for the excitonic condensate undergoing the dark-bright Josephson-type oscillations   \cite{Combescot1}. 
More recently it was suggested that the dark excitons can be brightened by imposing a strain on the monolayers of TMDs \cite{strain1, strain2}.   
Furthermore, dark excitons are normally characterized by energy smaller than the bright ones.
Thus, the formation of the bright excitonic order is obstructed by the conversion into the dark excitons as observed in Refs. \cite{Mysyr, Snoke}. It was not, however, possible to resolve the nature of the dark excitonic ensemble in these experiments, and no mechanism was also proposed for the fast speed of the conversion.

 Here we propose a method probing directly the coherence of the dark intervalley excitons. It is based on observing a significant modification of the polaritonic Rabi oscillations between a photon and a bright exciton in the presence of the dark excitonic condensate. The key element is the interaction with phonons responsible for the bright-dark interconversion. While in the absence of the dark condensate this interaction simply induces a decay of the Rabi oscillations, the nature of the oscillations changes dramatically once the dark condensate emerges.  The effect can be viewed as the formation of the polaritonic-type state between the phonon and the bright exciton which interferes with the polariton between the photon and the bright exciton. In short, we call this effect as photon-exciton-phonon polariton.
This proposal is quite generic regardless of the material. However, for the purpose of estimates we will be referring to the TMDs crystals.

\section{The minimal model}
The most relevant minimal terms in the Hamiltonian $H$ (cf. in 2D Ref. \cite{Ham}) describing conversion of a photon $\hat{p}, \hat{p}^\dagger$  into a bright exciton $\hat{b}, \hat{b}^\dagger$ and, then, to the dark intervalley one $\hat{d},\hat{d}^\dagger$ with the help of a phonon $\hat{c},\hat{c}^\dagger$ are
\begin{widetext}
\bea
H= \sum_{\bf k} \left[ \omega_p({\bf k}) p^\dagger_{\bf k} p_{\bf k} +   \omega_b({\bf k}) b^\dagger_{\bf k} b_{\bf k} + \omega_d({\bf k}) d^\dagger_{\bf k} d_{\bf k} + \omega_c({\bf k}) c^\dagger_{\bf k} c_{\bf k}\right] + H_{\rm int},
\label{H}
\eea
\end{widetext}
where the first four terms account for the free fields with the corresponding spectra $\omega_p({\bf k})$, $ \omega_b({\bf k}),\, \omega_d({\bf k}) ,\, \omega_c({\bf k})$ of  photons, bright excitons, dark excitons and phonons, respectively, in terms of their creation-annihilation operators; units in which $\hbar =1$ are used here and below. We assume that the bright and the dark exciton valleys are shifted by some momentum which is much larger than the typical monentum of a photon. The last term in (\ref{H}),  $H_{\rm int}=H_{pb} + H_{bdc}$, describes the optical dipole interaction  
\bea
H_{pb}= - \sum_{\bf k}  g_p(b^\dagger_{\bf k} p_{\bf k} +H.c.) 
\label{Hpb}
\eea
between photons and bright excitons, with $g_p$ determined by the dipole transition matrix element (directions are not shown)  and the coherence factors 
of the photons;
 the processes $H_{bdc}$ of the phonon assisted conversion between bright and dark excitons are accounted for by the term
\bea
H_{bdc}= -  \sum_{{\bf k}, {\bf q}} \frac{g_c\sqrt{v_0}}{\sqrt{V}}\left[b^\dagger_{\bf k} d_{\bf k -q} (c_{\bf q}+ c^\dagger_{-\bf q}) +H.c. \right],
\label{Hbdc}
\eea
where $V$ denotes a sample total volume and $v_0$ stands for the volume of the crystal unit cell. In 3D $v_0=l_0^3$ (and $v_0=l_0^2$ in 2D), with $l_0$ standing for a typical length $\sim 2-3 \AA$ of the unit cell. Here we have ignored a geometrical structure of the phonon matrix elements. 

Eq.(\ref{Hpb}) is written in the Rotating Wave Approximation (RWA). Its validity is limited by the condition $|g_p| << \omega_{p,b}$.  In Eq.(\ref{Hbdc}) $g_c$ stands for the effective interaction constant, with the factor $\sqrt{v_o}$ introduced for a convenience of the follow up discussion. The factor $\sqrt{V}$ guarantees that the expression (\ref{Hbdc}) results in the free energy which is extensive for $g_c=const$. The Hamiltonian (\ref{H},\ref{Hpb},\ref{Hbdc}) conserves the total number of photons and excitons $N_{ex}=\sum_{\bf k}  [p^\dagger_{\bf k} p_{\bf k} +  b^\dagger_{\bf k} b_{\bf k} + d^\dagger_{\bf k} d_{\bf k} ]$. Thus,
it is possible to introduce the chemical potential $\mu$ as $H \to H - \mu N_{ex}$ and set $\mu=\omega_d({\bf k}={\bf q}_0) $. Accordingly, in what follows $ \omega_p \to \omega_p - \omega_d({\bf q}_0)$ and $\omega_b \to \omega_b - \omega_d({\bf q}_0)$.

The term $\sim g_p$ in Eq.(\ref{Hpb}) is responsible for the celebrated polaritonic effect resulting in the superposition of photons and bright excitions leading to the polaritonic Rabi oscillations. In the absence of the dark condensate, the term (\ref{Hbdc}) leads to the decoherence of these oscillations by adding the factor $\exp(-\gamma t)$ with $\gamma \sim g_c^2$ (to be discussed below in more detail). The situation changes dramatically when the condensate of dark excitons forms. Here we, first, outline the main idea, and, then, develop it in more detail. In the case when a macroscopic number $N_d$ of dark excitons condenses at the momentum ${\bf q}_0$, the contribution from a single harmonic must be considered separately from all other harmonics in the sum (\ref{Hbdc}).  [This approach is in line with the traditional treatment of the condensate albeit at zero momentum---as described in Ref. \cite{LL}]. Indeed, in this case the operator  $d_{{\bf q}_0}$ can be treated classically as $d \to \sqrt{N_d}$, and the corresponding contribution to $H_{bdc}$ in Eq.(\ref{Hbdc}) becomes $H_d=  -  g_c \sqrt{n_d} \left[b^\dagger_0 (c_{{\bf q}_0}+ c^\dagger_{{-\bf q}_0}) +H.c. \right]$, with $n_d =N_dv_0/V$ determining the number of the condensed dark excitons per unit cell. Thus, at finite density of the condensate $n_d$ this harmonic dominates every harmonic corresponding to the excitations carrying momenta different from ${\bf q}_0$ and which scale as $\sim 1/\sqrt{V}$. Obviously, the question  about the decoherence due to the ensemble of these harmonics remains, and it will be addressed later. 
Below we will provide a more detailed derivation of the condensate part of the Hamiltonian by taking into account the existence of several valleys consistent with the point group symmetry.

\subsection{The dark condensate contribution.}
In the presence of the dark condensate in the valleys connected by the point group symmetry the dark exciton operator $\psi({\bf r}) $ acquires the c-number contributions associated with each such valley characterized by the momentum ${\bf q}_\alpha$, $\alpha=1,2, ...N_v$, such that $|{\bf q}_\alpha|=q_0$, with $N_v$ stands for the total number of the elements of the group.
Thus, in general, the condensate contribution to the second quantized excitonic operator $\psi =\frac{1}{\sqrt{ V}} \sum_{\bf k} d_{\bf k}\exp(i {\bf k} {\bf r})$ can be represented as
\be
\psi =\sum^{N_v}_{\alpha=1} \psi_\alpha  e^{i{\bf q}_\alpha {\bf r}} +\psi', \,\,  \psi_\alpha =\sqrt{N_\alpha/V}  e^{i \phi_\alpha},
\label{pgroup}
\ee      
where the replacement $d_{{\bf q}_\alpha} \to \sqrt{N_\alpha} e^{i \phi_\alpha}$ is made, with $N_\alpha, \phi_\alpha$ denoting, respectively, a  number of the condensed dark excitons and the phase in the $\alpha$-th valley; $\psi'$ accounts for the non-condensed excitons.

It is important to realize that the dark  (momentum-forbidden) excitonic condensate is not characterized by the lowest order coupling between phases $\phi_\alpha$ from different valleys---because of the momentum conservation requirement.  Indeed, each $\alpha$-component carries momentum ${\bf q}_\alpha$ which determines the spatial  fast oscillating factor $\sim e^{ i {\bf q}_\alpha {\bf r}}$ in Eq.(\ref{pgroup}). This precludes the traditional Josephson type coupling $\sim \psi^*_\alpha \psi_{\alpha'} +c.c.$ between different components, $\alpha \neq \alpha'$. However,  higher order Josephson-type couplings characterized by the terms of the energy $\sim \psi^*_\alpha \psi^*_{\bar \alpha} \psi_{\alpha'} \psi_{\bar \alpha'} \sim \cos(\phi_\alpha + \phi_{\bar \alpha} -\phi_{\alpha'} - \phi_{\bar \alpha'}) $ where $\alpha$ and $\bar{\alpha}$ refer to the momenta ${\bf q}_\alpha$ and $- {\bf q}_\alpha$, respectively, are not excluded. This aspect affecting the structure of the resulting Gross-Pitaevskii equation is promising of exotic dark excitonic phases which will be discussed elsewhere. 

 As usual in the physics of the superfluidity, it is reasonable to ignore the non-condensed portion $\psi'$ to the first approximation at low temperature $T$ (see in Ref.\cite{LL}).  Then, the dominant contribution from the dark excitons to the term (\ref{Hbdc})
becomes
\bea
H_{bdc}= - \sum^{N_v}_{\alpha} g_c\left[ \sqrt{n_\alpha} e^{i\phi_\alpha}b^\dagger_0 (c_{\alpha}+ c^\dagger_{\alpha}) +H.c. \right],
\label{Hbdc2}
\eea
where $n_\alpha=N_\alpha v_0/V$ and $c_\alpha, c^\dagger_\alpha$ are the phonon operators assisting the transitions between a bright exciton and a dark one characterized by the momentum ${\bf q}_\alpha$. Here we took into account  a significant difference between the momenta of the photons and bright excitons and $|{\bf q}_0|$ which allowed to set ${\bf k} \to 0$ in the operators $ b^\dagger_{\bf k} , b_{\bf k}$ in Eqs.(\ref{Hpb},\ref{Hbdc},\ref{Hbdc2}). 
Here we will consider the case of weak interactions between phonons and the excitons.  Then, the RWA approximation for the exciton-phonon interaction can be used, that is, $ b^\dagger_0 (\tilde{c}  +\tilde{c}^\dagger) +H.c. \to b^\dagger_0\tilde{c} +H.c.$  in Eq.(\ref{Hbdc3}).

Since the condensate $\psi_\alpha$ should be treated as a frozen parameter to the lowest approximation ignoring the normal excitations (see in Ref. \cite{LL}), the term (\ref{Hbdc2}) represents now a direct entanglement between bright excitons and the phonons leading to the polaritonic-type effect of the same nature as the one described by the term (\ref{Hpb}).

It is convenient to consider the orthonormal basis of complex  unit vectors  $e^{n}_\alpha$ (that is, $\sum^{N_v}_\alpha e^{n*}_\alpha e^{m}_\alpha=\delta_{mn}$) in the space of $N_v$ complex variables such that
\be
e^{1}_\alpha= \frac{1}{\sqrt{n_c}} \psi_\alpha, \quad  n_c=\sum_\alpha |\psi_\alpha|^2 .
\label{e1}
\ee 
 Then, it is straightforward to realize that only the superposition of the phonons 
\be
\tilde{c}_1 =\sum_\alpha e^{1*}_\alpha c_\alpha =\frac{1}{\sqrt{n_c}} \sum_\alpha \psi^*_\alpha c_\alpha 
\label{ac}
\ee  
is responsible for the coherent bright-dark interconversion, while the other superpositions $\tilde{c}_n= \sum_\alpha e^{n*}_\alpha c_\alpha$ with $n=2,3,... N_v$ are not (and participate only in the decoherence). At this point it is worth mentioning that the free phonon part of the Hamiltonian $H_{ph}=\sum_\alpha \omega({\bf q}_\alpha) c^\dagger_\alpha c_\alpha$ respects  the point group symmetry as well, with  
 $\omega({\bf q}_1)=\omega({\bf q}_2)= ...=\omega({\bf q}_{N_v})=\omega_c$ . Accordingly, the term (\ref{Hbdc2}) with the free phonon part responsible for the coherent mixing can be rewritten as
\be
H_{bdc}= - g_c\sqrt{n_d} \left[b^\dagger_0 \tilde{c} +H.c. \right] + \omega_c \tilde{c}^\dagger \tilde{c} ,
\label{Hbdc3}
\ee
with $\tilde{c}=\tilde{c}_1$ and $n_d=n_c v_0$ being a mean total number of the condensed excitons per unit cell.

 Thus,  in the presence of the condensate the contribution to the Rabi oscillations is accounted for by a single harmonic as discussed above 
\begin{eqnarray}
H_{R}&=&\omega_p p^\dagger p + \omega_b b^\dagger b + \omega_c \tilde{c}^\dagger \tilde{c} 
\nonumber \\
&&-[(b^\dagger(g_p  p + \tilde{g}_c  \tilde{c}) + H.c.],
\label{H2}
\end{eqnarray}
where the notation $\tilde{g}_c=\sqrt{n_d} g_c$ is introduced.
The effective Hamiltonian (\ref{H2}) describes completely the photon-exciton-phonon polaritonic effect.   The role of the decoherence is considered below.

\subsection{Decoherence effects.}
Decoherence is produced mainly by phonon-exciton scattering when bright excitons acquire a large momentum and become dark \cite{PT}. If the energy of the dark exciton valley is below that of the bright one, the decoherence rate $\gamma $ is finite even at zero temperature $T$ \cite{PT}.  
Here we focus on the effect due to the term $\sim g_c$ responsible for the bright-dark interconversion induced by phonons with momenta different from ${\bf q}_\alpha$. The estimates provided below are based on the assumption of smallness of the exciton-phonon interaction so that the second order perturbation theory can be applied. The opposite limit leading to the polaronic effect  will be addressed in a separate publication.

Here we provide estimates for typical values of $\gamma \approx 3-5$ meV in monolayers and bulk TMDs \cite{decoh2}  [ $\gamma$  can be as small as $\sim 1$ meV in special cases  \cite{decoh}].  In the case when $\omega_d < \omega_b$, the second order of the perturbation theory gives 
\be
\gamma = N_v\int \frac{d^3q}{(2\pi)^3} g_c^2v_0 \delta\left(\omega_b - \omega_c({\bf q}) - \frac{({\bf q} - {\bf q}_0)^2}{2 m} \right)   
\label{PT}
\ee 
in 3D at $T=0$,
where $\omega_b$ is taken at ${\bf k}=0$ (and shifted by $\omega_d({\bf q}_0)$); the integration is limited to the vicinity of one valley, with $m$ standing for the excitonic mass which will be taken as the bare electron mass. 

 The requirement that the polaritonic term $\sim \tilde{g}_c$ in Eqs. (\ref{Hbdc}),(\ref{Hbdc2}) dominates the dynamics reads 
\be
g_c \sqrt{n_d} >> \gamma \to  n_d >> \frac{\gamma^2}{g_c^2} .
\label{gc}
\ee
The maximum value of $\gamma$ is achieved for the dispersionless phonons (cf. \cite{PT}). Then,  $\gamma$ can be estimated as
$\gamma \approx (N_v/\pi^2) g_c^2  v_0 m\sqrt{2\delta \omega m}$ for $\delta \omega =\omega_b - \omega_c >0$ (and $\gamma=0$ otherwise). Thus, the condition (\ref{gc}) in the 3D can be written as
\be
n_d >>\frac{N_v}{\sqrt{2} \pi^2} \gamma l^3_0 m \sqrt{m \delta \omega},
\label{gc2}
\ee
and similarly in 2D as
\be
n_d >>\frac{N_v}{2 \pi} \gamma l^2_0 m. 
\label{gc3}
\ee
 Choosing $\gamma \approx 5$meV, $1/ml_0^2 \approx 3$ eV (a typical band gap),  $N_v=6$ and typical dark-bright energy differences around $\delta \omega \approx 50$ meV \cite{dif}, the r.h.s. in Eqs.(\ref{gc2}), (\ref{gc3}) can be estimated as $n_d > n_\gamma \approx 10^{-4} - 10^{-3}$, respectively. At finite $T$ around $\omega_c$, the decoherence rate increases due to the bosonic factor of phonons \cite{PT}. For a given $\delta \omega$ a significant increase begins at $T\sim 500$K, which is irrelevant because the condensation temperature in 3D at the dark condensate density, say, 10$^{-3}$ (per unit cell) is about 300 K. In 2D the condensate, strictly speaking, does not exist. However, the algebraic correlations persist and become essentially of the long-range type as $T$ lowers below the Berezinskii-Kosterlitz-Thouless temperature $T_{\rm BKT} \approx n_d/(ml_0^2)$ which for the same fraction gives $T_{\rm BKT} \approx 30$K.

The presented estimates show that the phonon induced decoherence effects should not disrupt the discussed polaritonic effect for the excitonic densities and temperatures  shown above. However, it is clear that it is more realistically to avoid the decoherence in the bulk samples than in monolayers.

In what follows we will describe our proposal while ignoring the decoherence effects. In general, this means that  a typical frequency of the coherent oscillations $\Omega \approx \sqrt{g_p^2 +\tilde{g}_c^2}$ (see below) is much larger than $\gamma$, which sets the longest time scale for the period of coherent oscillations $\tau_\Omega=2\pi/\Omega$ as $\tau_\gamma \approx 2\pi /\gamma$. (For the systems mentioned above  $\tau_\gamma \sim 1-10$ps). The estimates provided above, Eqs.(\ref{gc}-\ref{gc3}), indicate that there is a significant range of the condensate densities $n_\gamma <n_d<1$ where the decoherence does not obscure the oscillations.
Indeed, in the less favorable case $g_p << \tilde{g}_c$, we find that $\tau_\Omega /\tau_0 =\sqrt{ n_\gamma /n_d}\sim 1-10^{-2} $ for $n_d$ respectively increasing from $n_d \approx n_\gamma \sim 10^{-4}-10^{-3}$ to the values close to $n_d \sim 1$. This places the corresponding typical period of the oscillations within the range of $0.01-1$ps. Such values of $\tau_\Omega$ determine the time scale in the graphs presented below.

\section{Photon-exciton-phonon polaritonic effect}

Here we  discuss the coherent oscillations as described by the Hamiltonian (\ref{H2}) and ignore the decoherence effects, that is, assume that the condition (\ref{gc}) holds. First, for the sake of completeness, let's consider the standard polaritonic Rabi oscillations, and, then, analyze the full system. 

\subsection{Photon-exciton polaritonic effect}
The excitonc Rabi oscillations have been first observed in the fluorescence from indirect excitons in AgBr in magnetic field in Ref.\cite{10_Langer}, albeit with quite fast decay.  The time resolved oscillations of light emission from polaritons in quantum wells GaAs/Al$_{1-x}$Ga$_x$ was studied experimentally in Ref.\cite{4_Norris} and explained theoretically in Ref.\cite{11_Savona}. More recently, a much more distinct visibility has been achieved in the compound In$_{0.04}$Ga$_{0.96}$As in Ref
\cite{1_Dominici}.
Here we, first, describe the standard polaritonic oscillations ($\tilde{g}_c =0$ in (\ref{H2})), and, then will discuss their modification at finite $\tilde{g}_c$. 

 The spectrum and the eigenstates respectively are $\omega_\pm = \frac{\omega_p + \omega_b  \pm \sqrt{(\omega_b-\omega_p)^2 +4g_p^2}}{2} $ and
\beq
|\omega_\pm\rangle=\frac{1}{\sqrt{1+ (\omega_p - \omega_\pm )^2/g^2_p}}
\begin{pmatrix}
1  \\
(\omega_p - \omega_\pm)/g_p \\
\end{pmatrix}.
\label{S}
\eeq
An arbitrary state $|t\rangle$ at time $t$ can be expanded as $|t\rangle =C_+ |\omega_+\rangle e^{-i\omega_+ t} + C_- |\omega_-\rangle e^{-i\omega_- t}$ with $|C_+|^2 + |C_-|^2=1$, with the
initial condition 
$|t=0\rangle= \begin{pmatrix}
1 \\
0 \\
\end{pmatrix}$$=C_+ |\omega_+\rangle + C_- |\omega_-\rangle $ corresponding to one photon at $t=0$. The probability to observe the photon at $t>0$ becomes ${\cal P}(t)= |\langle t=0|t\rangle|^2$, that is,
\beq
{\cal P}(t)=1 - 2 |C_+|^2 |C_-|^2 \left[1- \cos[(\omega_+ -\omega_-)t]\right],
\label{P}
\eeq
 where $ C_+= \frac{\omega_- -\omega_p}{\omega_- - \omega_+}\sqrt{1+ (\omega_p - \omega_+ )^2/g^2_p}$, $C_-= \frac{\omega_p -\omega_+}{\omega_- - \omega_+}\sqrt{1+ (\omega_p - \omega_- )^2/g^2_p}$.
 Rabi oscillations,Eq.(\ref{P}),  take a very simple form at the resonance, $\omega_p=\omega_b$, as
$ {\cal P}(t)=\left[1 + \cos(2g_p t) \right]/2$. 

\subsection{Rabi oscillations at finite $n_d$} 
As long as  the dark condensate is formed,  the number of the Rabi frequencies tripple, and this leads to a wide variety of patterns in ${\cal P}(t)$ which are sensitive to the condensate density $n_d$.

At finite $n_d$, the Hamiltonian (\ref{H2}) can be represented as the 3$\times$3 matrix in the Hilbert space 
$|s\rangle= \begin{pmatrix} p \\ b \\ c \end{pmatrix} $, where $p,b,c$ represent, respectively, a photon, a bright exciton and a phonon. Thus, the spectrum is determined by
\beq
\begin{vmatrix} \omega - \omega_p &  g_p & 0 \\
g_p & \omega - \omega_b &  \tilde{g}_c \\
0 & \tilde{g}_c & \omega -\omega_c \end{vmatrix} =0,
\label{det2}
\eeq
or
\beq
(\omega -\omega_p)\left[ (\omega - \omega_b)(\omega -\omega_c) - \tilde{g}_c^2\right] - g_p^2(\omega - \omega_c)=0,
\label{spec2}
\eeq
with three eigenstates
\beq
|\omega \rangle=\frac{1}{\sqrt{1 + A^2 +B^2}} \begin{pmatrix} 1 \\ A \\ B \end{pmatrix},
\label{eig3}
\eeq
where $\omega=\omega_{1,2,3}$ are the roots of Eq.(\ref{spec2}) and $A=\frac{\omega_p - \omega}{g_p}$, $B= \frac{\tilde{g}_c (\omega_p -\omega)}{g_p (\omega_c-\omega)}$. Using the initial condition corresponding to one photon at $t=0$, that is, $|t=0\rangle=\begin{pmatrix}
1 \\
0 \\
0
\end{pmatrix} $, the probability ${\cal P}(t)$ becomes
\beq
{\cal P}(t)=1 - 2\sum_{i>j} |C_i|^2 |C_j|^2 \left[1- \cos[(\omega_i -\omega_j)t]\right] ,
\label{P2}
\eeq
where $C_i, i=1,2,3$ are the expansion coefficients in $|t\rangle =\sum_i C_i |\omega_i\rangle$ satisfying $\sum_i |C_i|^2=1$, 
$\sum_i C_i |\omega_i\rangle =|t=0\rangle$.

Let's start from the case $\omega_p=\omega_b=\omega_c$. [We remind that $\omega_p, \omega_b$ are counted from $\omega_d $ at the momentum ${\bf q}_0$]. 
In this case, the  eigenenergies and eigenstates, respectively, are $\omega_1=\omega_p, \omega_2= \omega_p+\Delta, \omega_3=\omega_p -\Delta$, with $\Delta =\sqrt{g_p^2 +\tilde{g}_c^2}$ and
\beq
|1\rangle =\frac{1}{\Delta}
\begin{pmatrix}
\tilde{g}_c  \\
0 \\
-g_p
\end{pmatrix}, \,\, |2\rangle =\frac{1}{\sqrt{2}\Delta}\begin{pmatrix}
g_p  \\
-\Delta \\
\tilde{g}_c
\end{pmatrix},   \,\, |3\rangle =\frac{1}{\sqrt{2}\Delta}\begin{pmatrix}
g_p  \\
\Delta \\
\tilde{g}_c
\end{pmatrix},
\label{Test}
\eeq 
Then, the probability ${\cal P}$ becomes as
\beq
{\cal P}(t)=\frac{\left| \tilde{g}_c^2 +g_p^2 \cos(\Delta t) \right|^2}{\Delta^4}.
\label{testP}
\eeq
The maxima ${\cal P}=1$ occur at $\Delta t_n =2\pi n, n=0,1,2,...$ and the minima ${\cal P}=\left(\frac{\tilde{g}^2_c - g^2_p}{\tilde{g}^2_c + g^2_p}\right)^2$ at $\Delta t_n =\pi (2n+1), n=0,1,2,...$ in the case $|\tilde{g}_c| > |g_p|$. In the opposite limit the minima ${\cal P}=0$ occur at $\Delta t_n =\arccos( - \tilde{g}_c^2/g_p^2) +2\pi n$.
Eq.(\ref{testP}) gives a way to measure $n_d$. 
\begin{figure}[!htb]
\vskip-8mm
\includegraphics[width=0.95 \columnwidth]{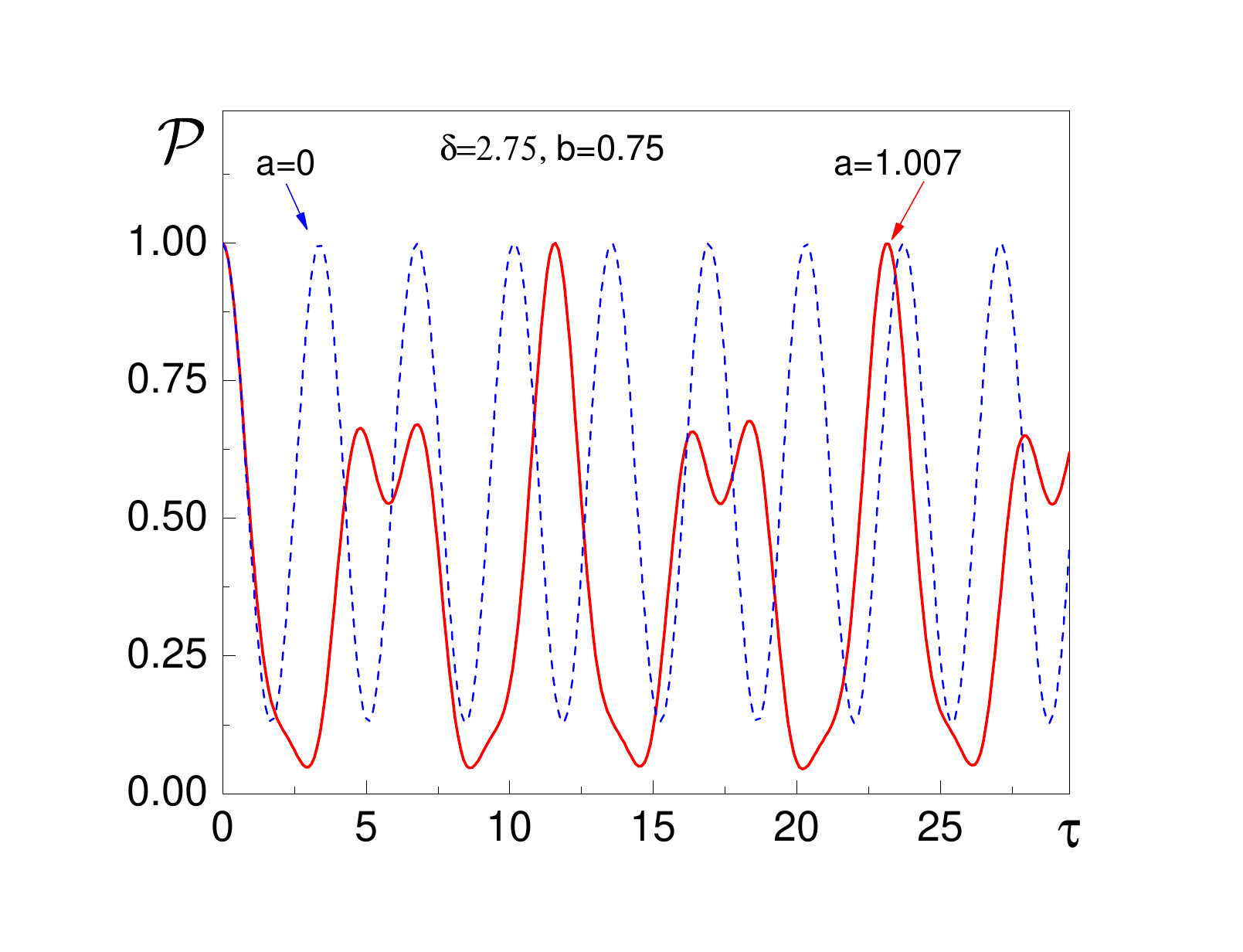}
	\vskip-8mm
\caption{ ${\cal P}(\tau)$ for the parameters shown close to the graphs. The dashed line represents oscillations at $a=0$ which correspond to the standard polaritonic Rabi oscillations. The solid line describes the oscillations at finite density $n_d$. } 
\label{fig0}
\end{figure}

\subsection{Quantum beats.}
It is convenient to consider the following parametrization:  $\omega_c=\omega_0 + \delta_c $, where $\omega_0$ coincides with one of the eigenfrequencies in the case $g_c=0$, say,
\beq
\omega_c =\frac{\omega_p + \omega_b  - \sqrt{(\omega_p-\omega_b)^2 +4g_p^2}}{2} +\delta_c , 
\label{omega0}
\eeq  
and $\delta_c $ stands for a deviation from this reference point.
Introducing the dimensionless quantities $z,y,\tau$ as $z=(\omega_p -\omega_b)/(\omega_p -\omega_c)$, $\omega= (2\omega_p - \omega_b -\omega_c)(y-1/3) + \omega_p$ and $\tau =(2\omega_p - \omega_b -\omega_c)t$, Eq.(\ref{spec2}) becomes
\bea
y^3 +py&& +q=0,
\label{y} \\
p= - \frac{1}{3} - a -b + \frac{z}{(1+z)^2},&& \,q= -\frac{1}{27} -\frac{p}{3} - \frac{b}{1+z},
\nonumber \\
z=-1 + &&\frac{4+2 \delta}{1\pm \sqrt{1+ 4b \frac{2+\delta}{1+\delta}}}, \label{z}
\eea
 where 
\bea
b=\frac{g_p^2}{(2\omega_p - \omega_b -\omega_c)^2},\, a=\frac{g_c^2}{g_p^2}bn_d ,\, \delta =\frac{\delta_c}{\omega_p -\omega_c}.
\label{abd}
\eea  
These three dimensionless parameters fully define all the options. Unless otherwise stated, in what follows we will be using  "+" in the denominator of Eq.(\ref{z}). 
\begin{figure}[!htb]
\includegraphics[width=1 \columnwidth]{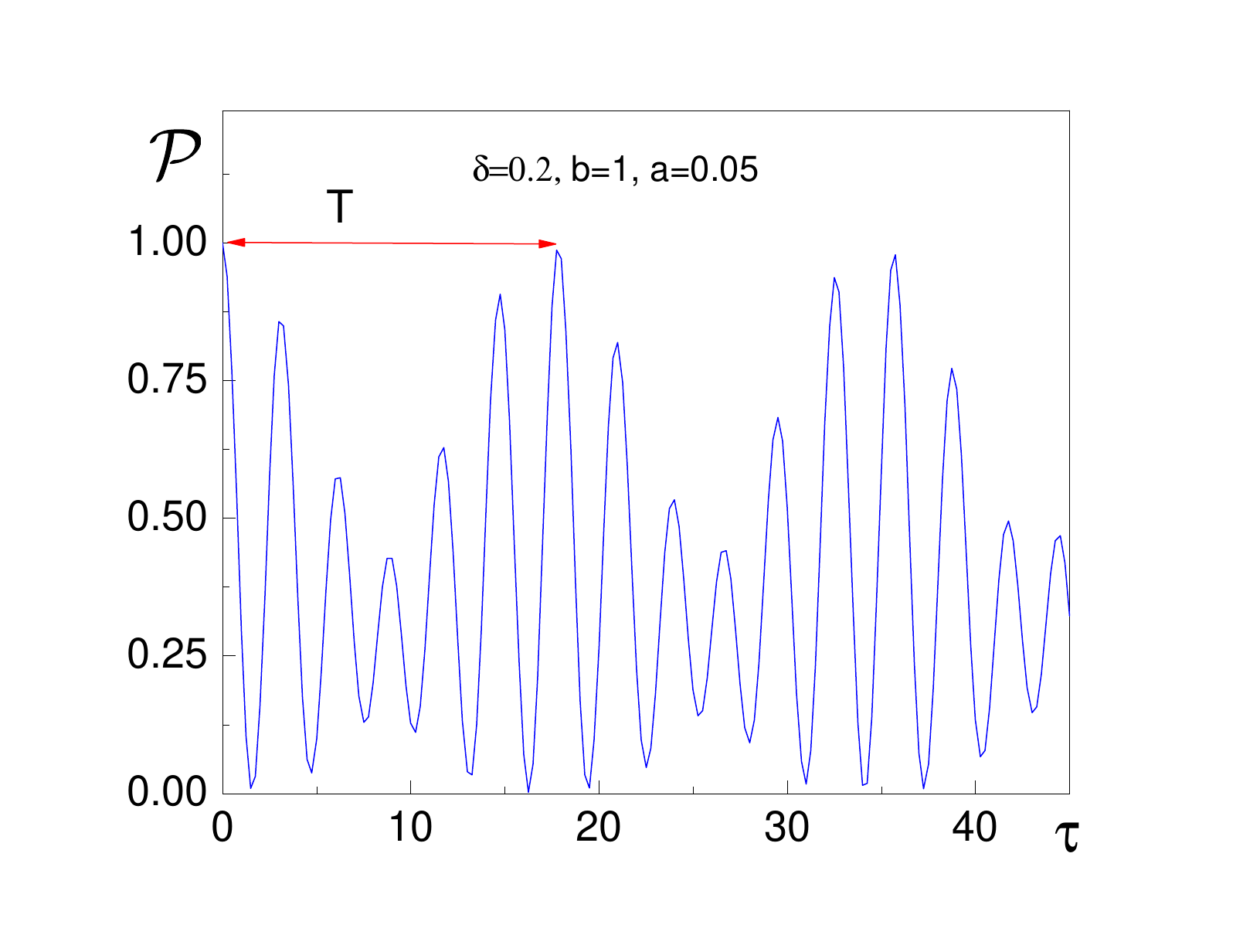}
\caption{The modulation of the oscillations for the parameters shown above the plot; $a_0=1.531$. }
\label{fig2}
\end{figure}
\begin{figure}[!htb]
\includegraphics[width=1\columnwidth]{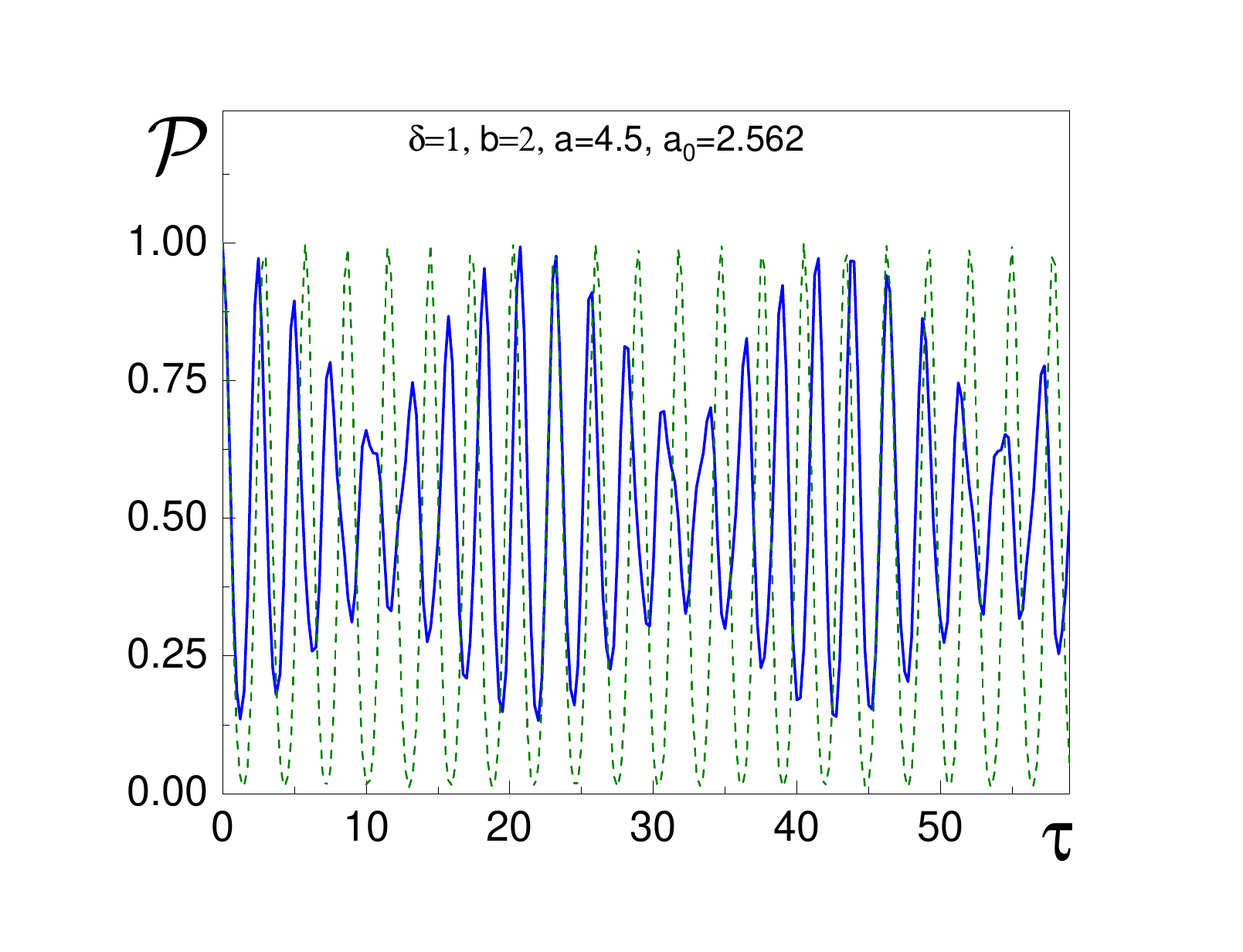}
\caption{The modulation of the oscillations for the parameters shown above the plot. The solid line corresponds to $a=4.5$; dashed line represents oscillations for $a=a_0$.  }
\label{fig2a}
\end{figure}
\begin{figure}[!htb]
\includegraphics[width=1 \columnwidth]{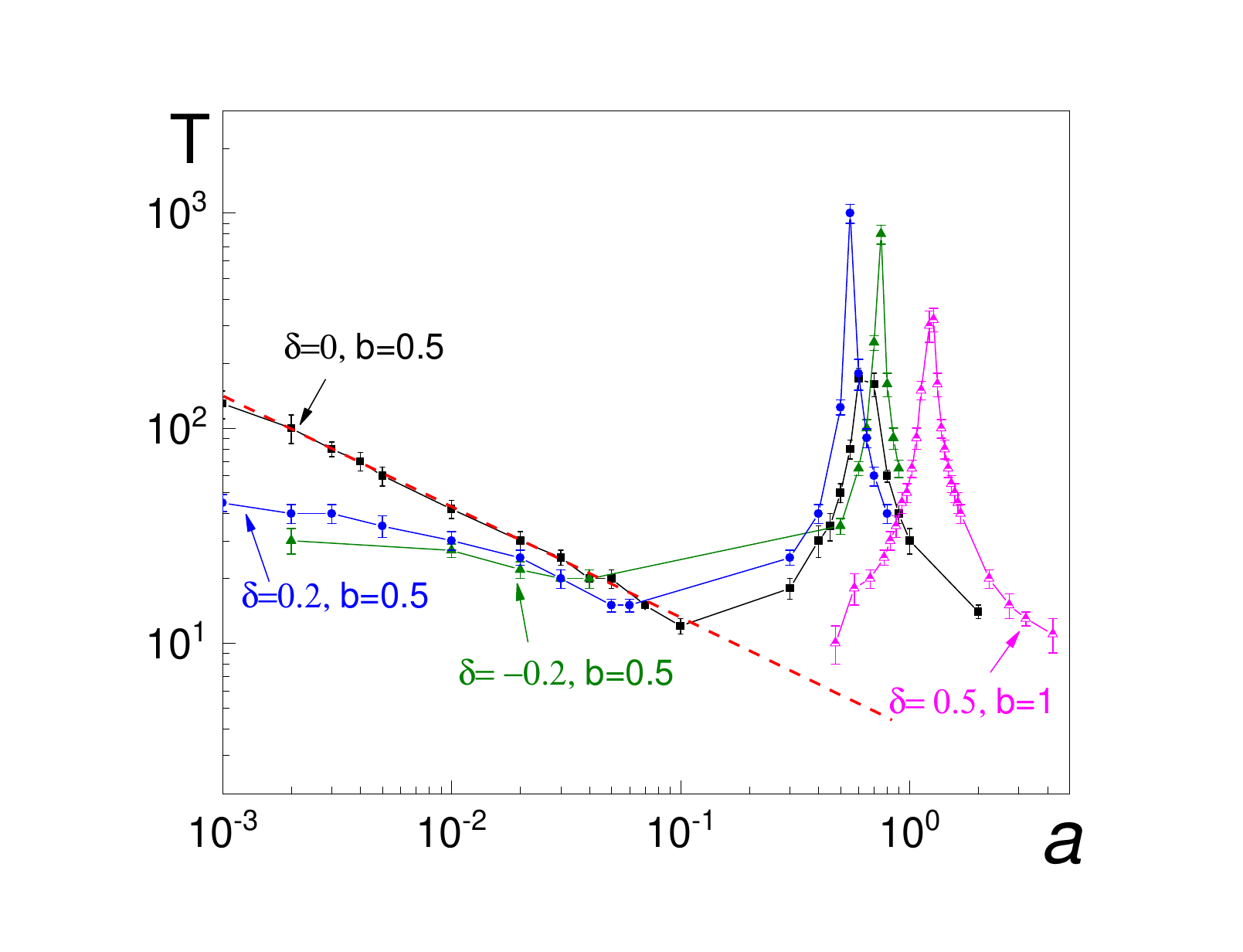}
\caption{ The period of the longest time moduation as a function of the parameter $a \sim n_d$ for several values of $b$ and $\delta$. The singularities occur at the points where $a=a_0$. }
\label{fig3}
\end{figure}
\begin{figure}[!htb]
\includegraphics[width=1 \columnwidth]{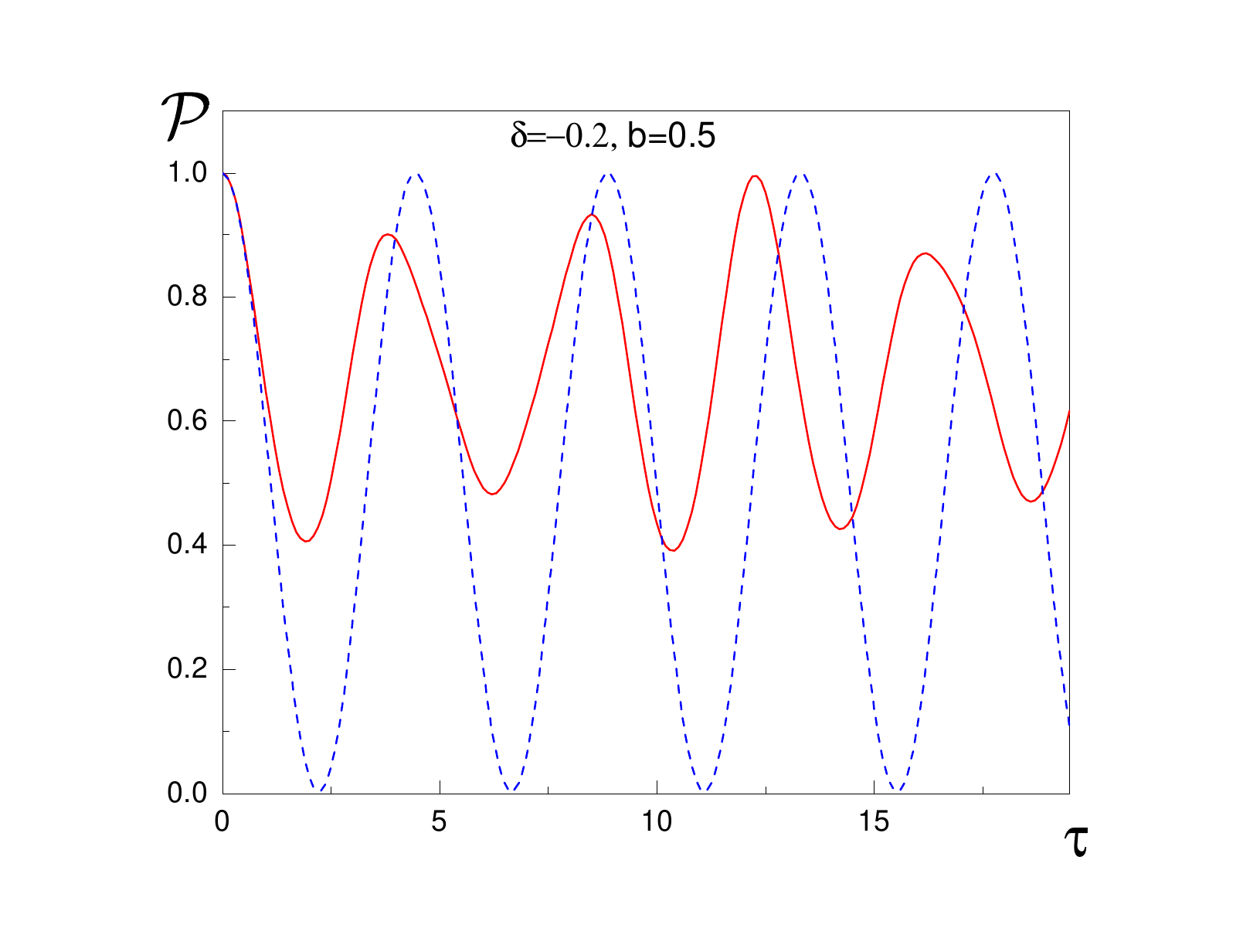}
\caption{ Rabi oscillations at finite $n_d$ (solid line for $a=2.5$, that is, $n_d/=0$) and at $a=0$ (dashed line). }
\label{fig4}
\end{figure}
\begin{figure}[!htb]
\includegraphics[width=1 \columnwidth]{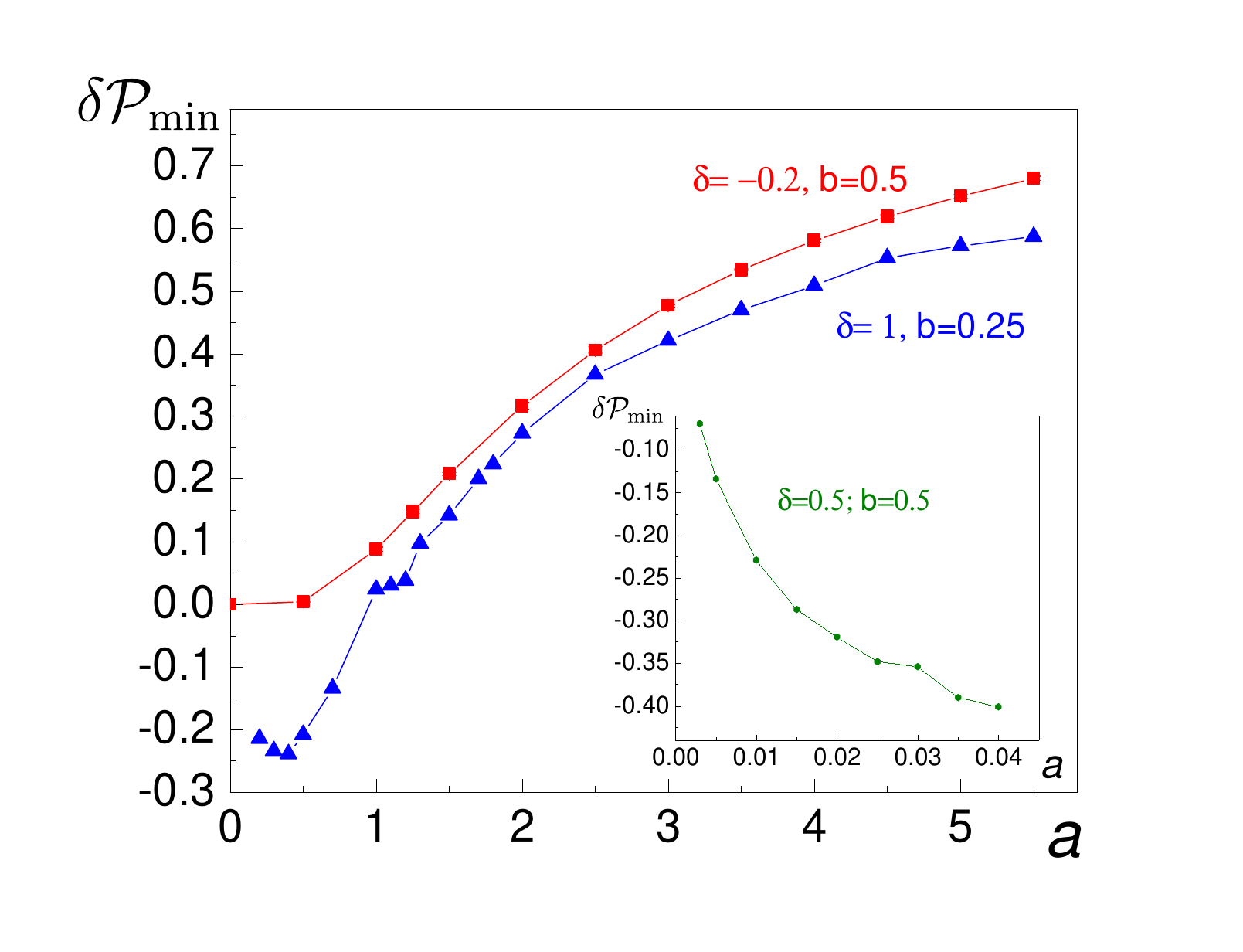}
\caption{ The difference $\delta {\cal P}_{\rm min}$ versus $a \sim n_d$ for the values $b,\delta$ shown close to each plot for the sign "+" in Eq.(\ref{z}). Inset: the data  for the sign "-" in Eq.(\ref{z}).}
\label{fig5}
\end{figure}
In general, various time patterns resulting from the three combinations $|\omega_i -\omega_j|$ can be observed.
One example is shown in Fig.~\ref{fig0}.  Such a qualitative difference between the cases $n_d=0$ (that is, $a=0$) and $n_d \neq 0$ (that is, $a\neq 0$)
can serve as an indication for the presence of the dark excitonic condensate. Extracting the values of $n_d$ can be done from several
measurements and, then, by fitting the resulting patterns by the parameters of the  above model.

It is instructive to consider a feature occurring  due to the existence of  points in the space of the parameters $a, b, \delta$ where one root of Eq.(\ref{y}) is zero (and the other two roots become $y_\pm =\pm \sqrt{-p}$). This corresponds to the condition $q=0$, that is,
\beq
a =a_0 = - \frac{2}{9} +\frac{z}{(1+z)^2} + \frac{2-z}{1+z}b ,
\label{a0}
\eeq
provided $a_0>0$. At this point there are two frequencies only,$\sqrt{-p}$ and $2\sqrt{-p}$, contributing to ${\cal P}(t)$ dynamics, as long as $p<0$. Accordingly, the pattern
will be strictly periodic with the period $2\pi/\sqrt{-p}$ which is not distinctly different from the case $n_d=0$. However, once $a$ deviates from $a_0$, all three harmonics contribute and, depending on $|a-a_0|/\sqrt{-p}$, a modulation of the harmonics   of $\sqrt{-p}$ and $2\sqrt{-p}$ develops. An example of such modulations (beats) are shown in Figs.~\ref{fig2},\ref{fig2a} with the period $T$ of the modulation depending on the difference $|a-a_0|$
so that $T \to \infty$ as $a \to a_0$. Typical dependencies of $T$ on $a\sim n_d$ for various $\delta$ and $b$ are shown in Fig.~\ref{fig3}. It is worth noting that for $\delta=0$ (the data fit by the dashed line) the dependence on $n_d$ is characterized by $T^{-1} \sim \sqrt{n_d} |g_c|$, that is, by about four orders of magnitude of the variation of  the dark condensate   density $n_d$.

Another feature, which is sensitive to the dark condensate density, is the dependence on $n_d$ of the offset from ${\cal P}=0$ of the Rabi oscillations. Examples for this are represented in Figs.~\ref{fig2a}, \ref{fig4}.
The difference $\delta {\cal P}_{\rm min}$ between the minimal values of ${\cal P}_{\rm min}$ at $n_d=0$ and $n_d\neq 0$  versus $a\sim n_d$ is shown in Fig.~\ref{fig5} for three sets of $\delta, b$.

\section{Discussion.}

The proposed method for detecting the dark excitonic condensate relies on single-phonon processes inducing the conversion between bright and dark excitons. Here we considered a simplified situation when both excitons are singlet. 
However, as discussed in Refs.\cite{Efros}, phonons can also induce spin flips. Thus, in some situations phonons can be responsible for the conversion between singlet  bright and triplet dark excitons.

Practically, among the three dimensionless parameters $a,b, \delta$ introduced above, Eqs.(\ref{omega0}--\ref{abd}), there are two tuning "knobs"---the difference $\delta_p= \omega_p -\omega_b$  
 between the photon and the bright exciton energies (which can be changed by adjusting the resonance condition in the cavity), and the intensity of the incoming light which controls the dark-condensate density $n_d$. 
This allows varying the dimensionless parameters within a relatively large margins. 

In our simplified approach we have ignored the geometrical structure of the dipole matrix elements and also the  phonon matrix elements responsible for the dark-bright conversion. As a result, the quantum beats become sensitive only to the total condensate density. An open question is to what extent the photon polarization effects can be used as a tool to distinguish the components of the condensate, if the full geometrical structure of the matrix elements is taken into account. This aspect will be considered elsewhere.

Validity of the RWA with respect to the phonons is limited by the condition $\tilde{g}^2_c << \omega_b \omega_c$.
It is worth mentioning that this requirement does not contradict to the limits $a>>1$ or $b>>1$. Indeed, as Eq.(\ref{abd}) indicates,
 large values of $a,b$ are possible when, e.g.,  $\omega_p < \omega_b $ , while $\omega_c< \omega_p$ when $z\to -1$.
Once $\tilde{g}_c$ becomes comparable to $ \omega_b \omega_c$,  the RWA with respect to the phonons is violated because of the  polaronic effect.  
Then, the created bright exciton can lower significantly  its energy due to the polaronic shift. This can disrupt the Rabi oscillations and lead to the fast conversion into the dark exciton.  In this case, the full Hamiltonian (\ref{H}) must be used, and the numerical diagrammatic approach developed in Refs.\cite{polaron0,polaron1,polaron2} must be applied, which is a subject of a future work. Here we conjecture that such a polaronic effect has already been observed  in the experiments  Refs.\cite{Mysyr, Snoke}.

{\it Summary}.---We have proposed a method for detecting the dark excitonic condensate based on the strong modification of the standard polaritonic Rabi oscillations. In the presence of the dark condensate the phonons become entangled with the bright excitons leading to the polariton-type effect between the phonon and the bright exciton. This endows  Rabi oscillations with features dependent on the dark condensate density. In particular, such a feature is the modulation (beatings) of the polaritonic oscillations,   with the longest period strongly dependent on the condensate density. Another feature is the offset of the oscillations. The presented results relying on weak interaction with phonons can be naturally extended to the case of the strong interaction leading to the polaronic effect.

{\it Acknowledgments}.---This work was supported
by the National Science Foundation under the grant DMR-2335905. We thank Boris Svistunov for expressing interest in the results and the perspectives for future work.

\end{document}